\documentclass[aps,prb,twocolumn,groupedaddress]{revtex4}
\usepackage{graphicx}
\usepackage{color}
\begin{document}
\title{Spatially-correlated microstructure and superconductivity\\in polycrystalline Boron-doped diamond}
\author{F.~Dahlem,$^{1}$ P.~Achatz,$^{1}$ O. A.~Williams,$^{2,*}$ D. Araujo,$^{3}$ E.~Bustarret$^{1}$ and H. Courtois$^{1}$}
\affiliation{$^{1}$Institut N\'eel, CNRS and Universit\'e Joseph Fourier, 25 Avenue des Martyrs, 38042 Grenoble, France,}
\affiliation{$^{2}$IMEC vzw, Division IMOMEC, Wetenschapspark 1, B-3590 Diepenbeek, Belgium}
\altaffiliation{Now at Fraunhofer Institute for Applied Solid State Physics, Tullastrasse 72, 79108 Freiburg, Germany.}
\affiliation{$^{3}$Departamento Ciencia de los Materiales e IMyQI, Universidad de Cadiz, 11510 Puerto Real, Spain.}
\date{\today}

\begin{abstract}
Scanning tunneling spectroscopies are performed below 100~mK on polycrystalline Boron-doped diamond films characterized by Transmission Electron Microscopy and transport measurements. We demonstrate a strong correlation between the local superconductivity strength and the granular structure of the films. The study of the spectral shape, amplitude and temperature dependence of the superconductivity gap enables us to differentiate intrinsically superconducting grains that follow the BCS model, from grains showing a different behavior involving the superconducting proximity effect. 
\end{abstract}
\maketitle

Over the last few years, superconductivity has been discovered in heavily doped group IV covalent semiconductors,\cite{Blase375} in particular diamond \cite{Ekimov542} and silicon.\cite{bustarret465} In the case of diamond, low temperature superconductivity appears at the same doping level than the metallic state created by heavy Boron doping.\cite{klein165313} Evidence for a pairing mechanism mediated by phonons in the weak coupling limit has been provided among others by very low temperature scanning tunneling spectroscopy of single crystal epilayers.\cite{sacepe097006} Polycrystalline diamond films can be a new model system for the general issue of the nature of superconductivity in strongly disordered metals.\cite{Larkin-Ann} In such systems, disorder sits either at the atomic-scale, in which case electronic excitations can become localized so that superconductivity vanishes \cite{Escoffier-PRL} or at a larger scale, for instance that of a granular structure, in which case the two competing mechanisms are the Coulomb blockade and the superconducting proximity effect.\cite{Moussy-EPL,Belzig-PRB} Nevertheless, recent studies of polycrystalline diamond films \cite{Dubrovinskaia-PNAS,PRB-Willems} did not provide a clear picture on the co-existence between superconductivity and disorder in these films.

In this paper, we report a study of the local superconducting and structural properties of high-quality polycrystalline Boron-doped diamond by very low temperature Scanning Tunneling Microscopy (STM). The granular structure was consistently characterized by STM and Transmission Electron Microscopy (TEM). In contrast with epitaxial films, a strong correlation is observed between the granular microstructure and the superconductivity local strength. The spatial evolution and temperature dependence of the local electronic density of states are consistent with the picture of an assembly of grains, which either follow the BCS model or present another superconducting behavior involving the superconducting proximity effect. 

\begin{figure}[b]
\includegraphics[width=\columnwidth]{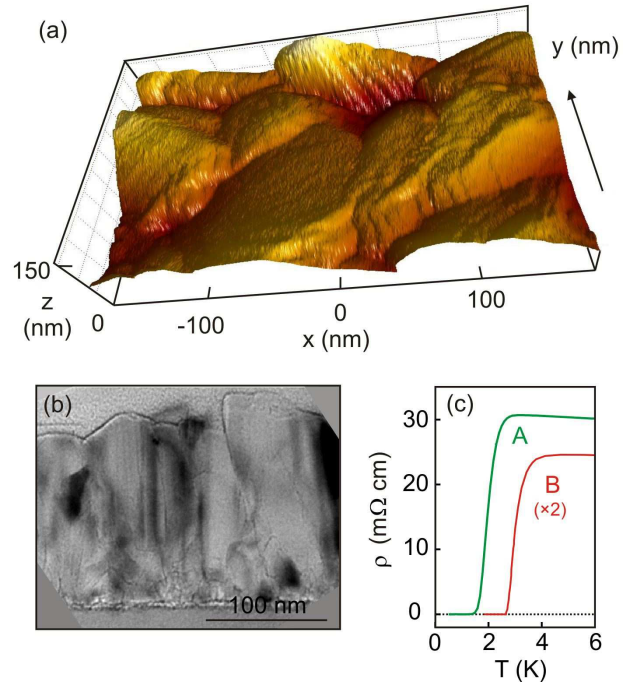}
\caption{(Color online)
a) 3D-view of a 350 x 128 nm$^2$ STM topography on the sample A Boron-doped polycrystalline diamond film showing a faceted granular morphology. 
b) TEM cross-section image in bright field condition of sample B.
c) Superconducting transition observed via the measurement of resistivity versus temperature for samples A and B.
\label{SamplePresentation}}
\end{figure}

Boron-doped polycrystalline diamond thin films of different thicknesses were grown as described elsewhere \cite{Williams-DRM,achatz201203} by microwave plasma-enhanced chemical vapor deposition from hydrogen-rich methane-trimethylborane-hydrogen gaseous mixtures on ultrasonically seeded quartz (sample A) and oxidized silicon (sample B) substrates. As shown by Fig.~\ref{SamplePresentation}a and b displaying respectively a very low temperature STM \cite{Moussy-RSI,Gupta-PRB} topography of sample A and a TEM cross section in bright field condition of sample B, the roughness of both films of the order of 80 nm is associated to well-defined facets. Beside confirming the origin of the large scale roughness, the TEM micrograph reveals a grain configuration with three different regions from bottom to top: (i) diamond seeds covering the substrate surface, (ii) a first layer nucleated on some of these seeds, with nearly equi-axial grains of a diameter below 50 nm and many grain boundaries almost perpendicular to the growth axis, (iii) and finally larger columnar grains with an average size of 150 nm, inducing the facetted aspect of the free surface. In the latter region, most grain boundaries, but not all, are oriented parallel to the growth axis. Selective area diffraction patterns (not shown here) confirm that in the intermediate region the grains are randomly oriented, while the top region has a stronger texture, with growth directions distributed between the (111) and (001) crystallographic orientations. On a given facet, the situation is thus close to that on an epilayer. During the growth, facets with a different orientation are expected to take up Boron with a different efficiency,\cite{Ushizawa19981719} inducing a local Boron concentration variation that may reach a factor up to 8.

Sample A was 1 $\mu$m-thick and had a normal state resistivity of 26 m$\Omega$.cm. Its macroscopic superconducting transition defined at half the normal state resistance occurred around 2.0 K with an onset at 2.9 K, see Fig.~\ref{SamplePresentation}c. This sample appears in Ref.~\onlinecite{achatz201203} under the label Bus10000. Sample B was 200 nm-thick and had a normal state resistivity of 12.3 m$\Omega$.cm. Its superconducting transition occurred around 3.0 K with an onset at 4.1 K, see Fig.~\ref{SamplePresentation}c. The critical temperature is thus, as expected, higher in the sample with the higher doping and lower resistivity, and the values are in line with those previously reported.\cite{APL-Nesladek} Sample B Boron concentration was measured by Secondary Ion Mass Spectroscopy to be 3.2 10$^{21}$ B/cm$^3$ at the surface and 3.5 10$^{21}$ B/cm$^3$ in the bulk. Although both samples are doped far above the insulator-to-metal transition found to occur at 2.5 10$^{20}$ B/cm$^3$ in similar films,\cite{Gajewski-PRB} they retain a resistivity one order of magnitude higher than that of single crystal epilayers with a similar Boron content, where the mean free path is about 1 nm.\cite{bustarret205} The grain boundaries thus add an important contribution to the resistivity of granular films. However, transport and magnetic susceptibility measurements have clearly demonstrated the non-filamentary nature of their superconductivity,\cite{achatz201203} as well as the good electrical coupling between the grains.\cite{Gajewski-PRB}

\begin{figure}[t]
\includegraphics[width=\columnwidth]{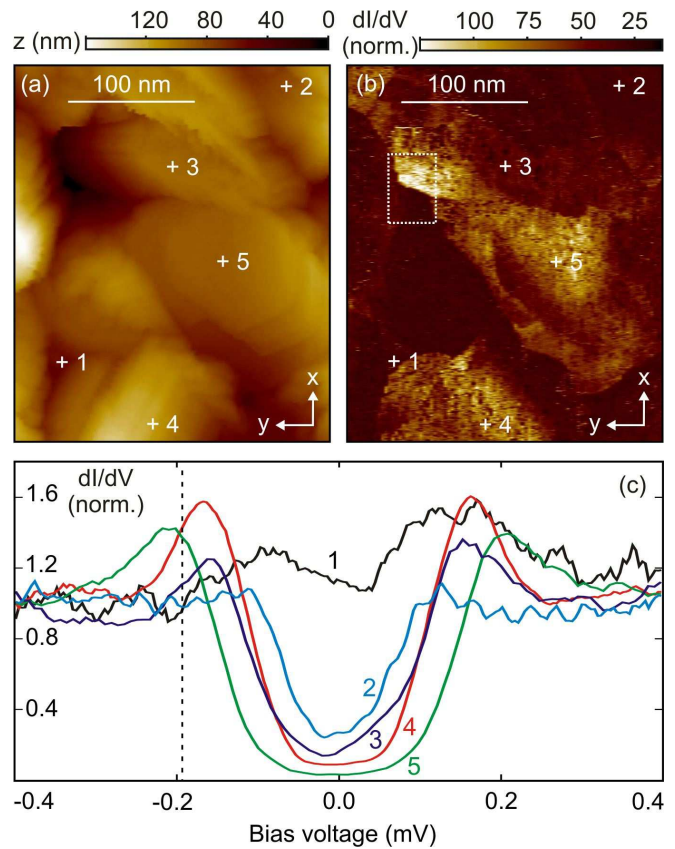}
\caption{(Color online)
a) 2D-view of Fig.~\ref{SamplePresentation}a Sample A image. The \mbox{d.c.} voltage bias (vertical dotted line in (c)) and the set-point current are - 0.19~mV and 250~pA respectively, giving a tunnel resistance of 0.76~M$\Omega$. 
b) Differential conductance $dI/dV$ image acquired simultaneously, thanks to an \mbox{a.c.} bias modulation of 25~$\mu$eV. Bright areas correspond to strongly superconducting grains with large coherence peaks, whereas dark areas refer to grains without coherence peak or with a metallic behavior. 
c) Local differential conductance spectra measured at a temperature below 100~mK at different locations indicated on images (a) and (b).
\label{TopoSpectro}}
\end{figure}

As for STM topography, a \mbox{d.c.} voltage bias was applied between a W tip and the sample while the tip was scanned in a constant current mode. Several STM runs without specific surface cleaning gave similar results with a good spatial resolution on both samples A and B. While making the image shown in Fig. \ref{SamplePresentation}a, a 25~$\mu$eV \mbox{a.c.} voltage at a 2~kHz frequency was added to the - 0.19 mV \mbox{d.c.} bias voltage. The related current modulation directly yielded the local differential conductance, which scales with the local density of states at an energy determined by the \mbox{d.c.} bias, smeared out by the thermal energy $k_BT$. We have checked that the \mbox{a.c.} current variation due to imperfections in the tunnel resistance regulation had a negligible contribution.

Fig.~\ref{TopoSpectro} top displays in gray scale the topography and the differential conductance images of the same region. As for the differential conductance (Fig.~\ref{TopoSpectro}b), bright areas indicate a large conductance and correspond to strongly superconducting grains with large coherence peaks (see spectra 4 and 5 in Fig.~\ref{TopoSpectro}c at - 0.19 mV). Dark areas indicate a lower conductance and refer to grains without coherence peak or with a metallic behavior (see spectra 1, 2 and 3 in Fig.~\ref{TopoSpectro}c at - 0.19 mV). The comparison between the two images reveals that, near the surface, the superconductivity is correlated to the granularity. This is the most important result of the present work. A similar behavior was observed in samples A and B over a series of experimental runs, on a series of locations over the millimetric-sized doped area and with different tips. The correlation of the measured spectra with the granular structure also indicates that our STM spectroscopies provide a picture of the electronic properties of the granular film, not of a possible contamination layer. Our Boron-doped diamond films are thus a mixture of grains or regions, connected to each other, but with different superconducting coupling strengths.  

Fig.~\ref{TopoSpectro}c presents the differential conductance probed at several positions indicated on Fig.~\ref{TopoSpectro}a or b. We have checked that tunneling spectra did not depend on the tunnel resistance set-point in the range 1-10 M$\Omega$. In contrast to early STM measurements on similar samples,\cite{nishizaki22,troyanovskyi27} we observe in many locations an almost fully opened gap, presumably thanks to a better sample surface morphology. Within one grain, the electronic properties change smoothly as expected for a slow variation of doping concentration (small black spots are noise spikes). At a grain boundary such as position 1, the spectrum is close to constant and therefore metallic-like. At positions 4 and 5, the spectrum displays an energy gap of the order of about $200~\mu$eV and two coherence peaks, characteristic of an intrinsic superconductor. On other grains, at position 2 or 3, the gap and the peaks are less pronounced, and the superconductivity is weaker. 

The interface between two types of grains may be locally opaque, as in the region of strong contrast highlighted by a rectangle in Fig.~\ref{TopoSpectro}b. Most of the junctions were found to be more transparent as can be seen following a line between positions 4 and 5, with progressive variations of the gray scale reminiscent of the superconducting proximity effect. We did not observe any quantum confinement effect \cite{PRL-Altfeder} within a grain, presumably because of the strong inter-grain coupling in our samples. Our results are in strong contrast with a recent STM study on one polycrystalline diamond film, where a strong and short-range modulation of the superconductivity was observed within the same grain.\cite{PRB-Willems} We believe such a modulation to result from an irregular and strongly disordered surface. Our samples feature a much smoother surface corrugation, which enabled us to combine simultaneously high quality topography and spectroscopy, and to observe effects that could be hidden otherwise.

\begin{figure}[t]
\includegraphics[width=\columnwidth]{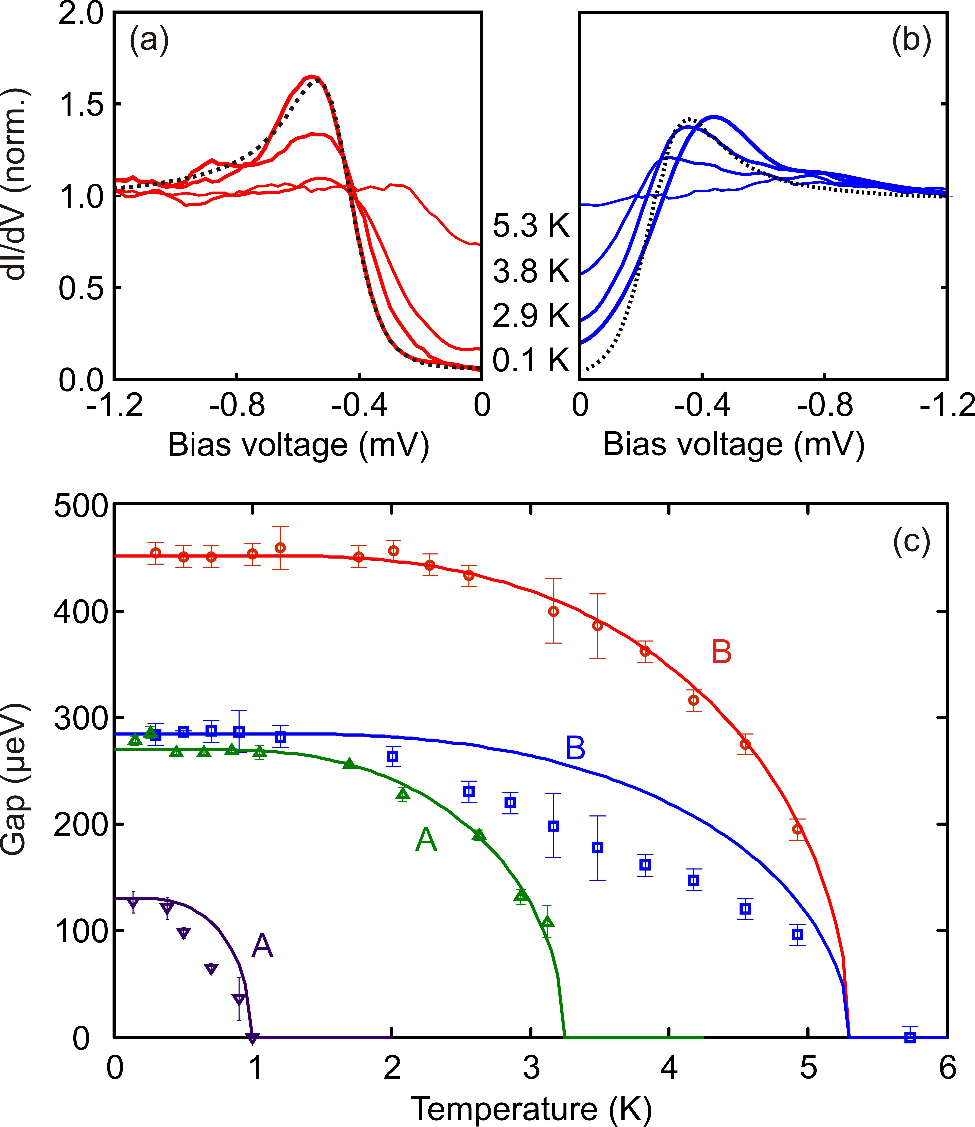}
\caption{(Color online)
a, b) Differential conductance spectra obtained on two different locations in sample B at selected temperatures, featuring a BCS (a) or a non-BCS behavior (b). The dotted lines are BCS fits of the lowest temperature spectra, taking into account an effective temperature of 600 mK.
c) Temperature dependence of the local energy gap measured at different locations on samples A and B. Sample B data correspond to Fig.~\ref{Gaps}a and b data, but over a broader temperature set. Continuous lines are fits to the BCS model.
\label{Gaps}}
\end{figure}

We measured the amplitude of the local superconducting energy gap as a function of temperature in a series of different locations. Two representative data sets obtained on sample B are displayed in Fig.~\ref{Gaps}a and b, showing a different kind of behavior. Fig.~\ref{Gaps}a curves can be correctly fit with a BCS-type equation (without any Dynes parameter), which defines the first kind of data. The very low temperature data were fit taking into account an effective electronic temperature of 600 mK. In contrast, the data from Fig. \ref{Gaps}b cannot be fit by the BCS equation, a behavior which defines the second kind of data. In this case, the value of the energy gap was deduced from the inflection points of the spectra. These fits provided us with the temperature dependence of the local superconducting energy gap $\Delta$. We have checked that, when both applicable, the two methods yield the same gap value.

Fig.~\ref{Gaps}c shows the temperature dependence of the local energy gap for the two sample B locations discussed above, and for two similar ones in sample A. As expected, the local energy gap always decreases when the temperature rises and it vanishes at a given temperature, identified to a local critical temperature. The temperature dependence at locations of the first kind can be well fit by the BCS prediction for the gap temperature dependence. In contrast, data sets of the second kind do not follow the BCS dependency, but rather a linear behavior. The local critical temperature of some of the individual grains in both samples appears above the respective macroscopic critical temperature (see Fig.~\ref{SamplePresentation}c). This behavior agrees with the observed significant width of the resistive transition, which is then related to the appearance of a percolating path through sufficiently well coupled superconducting grains with different gaps, as confirmed by susceptibility measurements.\cite{achatz201203} The local $\Delta (0)/k_BT_c$ ratio values were found to be significantly lower than the 1.76 value expected for a conventional BCS superconductor. This could be explained by the inverse proximity effect due to the contact with grains with a weaker superconducting coupling. 

As for the second kind of locations, the spectral shape, the small amplitude and the temperature dependence of the energy gap all indicate a superconductivity that is not conventional BCS. We ascribe the significantly weaker superconductivity in these regions to a locally lower doping, presumably related to different facet orientations. The related tunneling spectra are then affected by the proximity with neighboring strongly superconducting grains. Across the junction between a superconductor and a normal metal, the local density of states is known to evolve from a U-shaped BCS spectra to a V-shaped pseudo-gap spectra induced by proximity effect.\cite{Moussy-EPL,Belzig-PRB} In our experimental data, we precisely observe both U-shaped spectra with a large gap (see Fig.~\ref{Gaps}a) and V-shaped spectra with a smaller gap (see Fig.~\ref{Gaps}b). In the diffusive regime of relevance here, proximity effect occurs on a characteristic length $\xi_S = \sqrt{\hbar D/2\Delta(0)}$,\cite{Belzig-PRB} where $D$ is the diffusion constant. The characteristic scale for induced superconductivity appears here to be larger than the mean calculated value $\xi_S \simeq$ 1.3 nm, based on measured resistivities. This discrepancy can be understood by the fact that although a grain is intrinsically non-superconducting, non-zero electronic attractive coupling can reinforce the observed proximity effect.

In summary, superconducting polycrystalline diamond films can be described as a disordered network of superconducting grains coupled through transparent junctions. More precisely, our study demonstrates the intrinsic superconductivity of individual grains and the broad distribution of their superconducting gap values. This heterogeneity must be taken into account when designing innovative superconducting devices \cite{Mandal} taking advantage of the relatively high critical field of Boron-doped diamond. In contrast with earlier speculations,\cite{Dubrovinskaia-PNAS} the grain boundaries do not appear as a specifically favorable region for superconductivity. Our conclusions are consistent with a superconductor to insulator transition scenario driven by a competition between Coulomb blockade and superconducting proximity effect. Since disorder is here directly related to the material granular structure and thus directly accessible, Boron-doped diamond films can be considered as a model system for the local study of strongly disordered superconductors. 

The authors thank M.~P.~Alegre for TEM preparation, F.~Jomard for SIMS measurements, C.~Chapelier, T.~Dubouchet, S.~Mandal, D.~Roditchev for discussions. We acknowledge support from EU MICROKELVIN infrastructure grant 228464 and ANR grant 09-BLA-0170.

\end{document}